\documentclass[12pt]{article}
\usepackage{epsfig}

\newcommand{\mysection}{\setcounter{equation}{0}\section}
\renewcommand{\theequation}{\thesection.\arabic{equation}}
\def\beq{\begin{equation}}
\def\eeq{\end{equation}}
\def\beqa{\begin{eqnarray}}
\def\eeqa{\end{eqnarray}}
 
\newlength{\dinwidth} \newlength{\dinmargin}
\begin{document}

\begin{center}
{\Large \bf Two-loop soft anomalous dimensions for single top quark 
associated production with a $W^-$ or $H^-$}
\end{center}
\vspace{2mm}
\begin{center}
{\large Nikolaos Kidonakis}\\
\vspace{2mm}
{\it Kennesaw State University,  Physics \#1202,\\
1000 Chastain Rd., Kennesaw, GA 30144-5591, USA}
\end{center}
 
\begin{abstract}
I present results for the two-loop soft anomalous dimensions for   
associated production of a single top quark with a $W$ boson or a 
charged Higgs boson. The calculation uses expressions for the 
massive cusp anomalous dimension, which are presented in different forms, 
and it allows soft-gluon resummation at 
next-to-next-to-leading-logarithm (NNLL) accuracy.  
From the NNLL resummed cross section 
I derive approximate NNLO cross sections for $bg \rightarrow tW^-$ and 
$bg \rightarrow tH^-$  at LHC energies of 7, 10, and 14 TeV. 
\end{abstract}
 
\mysection{Introduction}
 
The Large Hadron Collider (LHC) will produce top quarks via top-antitop pair 
or single top quark processes with relatively large cross sections. Given the 
importance of the top quark \cite{toprev} to electroweak and Higgs physics, 
and the observation of single top events at the Tevatron \cite{D0st,CDFst,singletop}, 
it is crucial to have a good theoretical understanding of top quark production cross 
sections. An interesting channel to study is associated production of a top quark with a 
$W$ boson, $bg \rightarrow tW^-$, which is sensitive to new physics 
and allows a direct measurement of the $V_{tb}$ CKM matrix element. 
This process is very small at the Tevatron but has the second 
highest cross section among single top processes at the LHC. 
A related process is associated production of a top quark with a charged Higgs boson, 
$bg \rightarrow tH^-$. Charged Higgs bosons appear in the Minimal Supersymmetric
Standar Model (MSSM) and other two-Higgs-doublet models (2HDM).
In the MSSM there are two Higgs doublets, one giving mass to the 
up-type fermions and the other to the down-type fermions.
Among the extra Higgs particles in the MSSM are two charged Higgs bosons, 
$H^+$ and $H^-$, and the associated production of a top quark with a 
charged Higgs is a process that the LHC has good potential 
to observe.
Since a central mission of the LHC is to find the Higgs boson and another is to look for supersymmetry, 
the associated production of a charged Higgs with a top quark is an important 
channel to study.

The next-to-leading order (NLO) corrections to $bg \rightarrow tW^-$ were calculated in \cite{ZhutW} 
and to $bg \rightarrow tH^-$ in \cite{ZhutH,Plehn,BHJP}. 
These processes are very similar with respect to QCD corrections and they have the 
same color structure.
Soft-gluon emission is an important contributor to higher-order corrections, particularly 
near partonic threshold. The soft-gluon corrections can be formally resummed to all 
orders in perturbation theory. The resummation follows from the factorization of the 
cross section into a hard-scattering function $H$ and a soft function $S$ that describes 
noncollinear soft-gluon emission in the process \cite{NKGS,NKst}. 
The renormalization group evolution of the soft function  is controlled by a 
process-dependent soft anomalous dimension $\Gamma_S$. The calculation of $\Gamma_S$ 
is performed in the eikonal approximation, which describes the 
emission of soft gluons from partons in the hard scattering and leads to modified 
Feynman rules in diagram calculations.
At next-to-leading-logarithm (NLL) accuracy these corrections were resummed for 
$tW^-$ production at the Tevatron and at the LHC in \cite{NKst,NKstlhc,NKtop}, 
while the corrections for $tH^-$ production were presented in \cite{NKchiggs,NKchiggs2}. 
These results involved the calculations of the one-loop soft anomalous dimension for 
these processes.

Recent developments in two-loop calculations \cite{NK2l,DPF092l,NKsch} have now made possible 
the resummation of next-to-next-to-leading-logarithm (NNLL) corrections for QCD processes. 
Here we begin by calculating the two-loop soft (cusp) anomalous dimension for two massive quarks,  
and then using these results in the limit when one quark is massive (top quark)
and one is massless (bottom quark) we calculate the diagrams 
for associated single top quark production. Since there are three colored 
partons in the partonic processes $bg \rightarrow tW^-$ and $bg \rightarrow tH^-$ 
there are many diagrams to consider but the end result for the two-loop soft anomalous dimension for these processes can be written in a 
simple formula.
We then use those results to calculate approximate next-to-next-to-leading 
order (NNLO) cross sections for $tW^-$ and $tH^-$ production at the LHC.

\mysection{Two-loop soft (cusp) anomalous dimension for a heavy quark-antiquark pair}

We begin by presenting the calculation of the two-loop cusp anomalous dimension, which is 
the soft anomalous dimension for $e^+ e^- \rightarrow t {\bar t}$ 
\cite{NK2l,DPF092l}.

We expand the soft (cusp) anomalous dimension as
$\Gamma_S=(\alpha_s/\pi) \Gamma_S^{(1)}+(\alpha_s/\pi)^2 \Gamma_S^{(2)}+\cdots$,
The one-loop soft anomalous dimension, $\Gamma_S^{(1)}$, can be read
off the coefficient of the ultraviolet (UV) poles of the one-loop diagrams
in Fig. 1. 

In the eikonal approximation, as the gluon momentum goes to zero, the quark-gluon
vertex reduces to $g_s T_F^c \, v^{\mu} / v\cdot k$, 
with $g_s$ the strong coupling, $v$ a dimensionless velocity vector, 
$k$ the gluon momentum,
and $T_F^c$ the generators of SU(3) in the fundamental representation.
For example the integral for the diagram in Fig. 1(a) is given by
\beq
\frac{\alpha_s}{\pi} \, I_{1a} = g_s^2 \int\frac{d^n k}{(2\pi)^n} 
\frac{(-i)g_{\mu \nu}}{k^2} \frac{v_i^{\mu}}{v_i\cdot k} \, 
\frac{(-v_j^{\nu})}{(-v_j\cdot k)} 
\eeq
where $i$ labels the quark and $j$ the antiquark. 
The quark and antiquark velocity vectors obey the relations
$v_i \cdot v_j=(1+\beta^2)/2$ and 
$v_i^2=v_j^2=(1-\beta^2)/2$, where $\beta=\sqrt{1-4m^2/s}$ with
$m$ the heavy quark mass and $s$ the center-of-mass energy squared.
The eikonal diagrams are calculated in dimensional regularization with 
$n=4-\epsilon$ using Feynman gauge in momentum space. 

\begin{figure}
\begin{center}
\includegraphics[width=9cm]{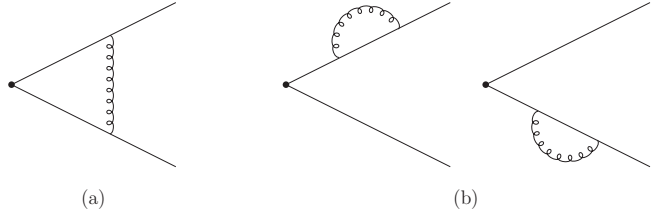}
\caption{One-loop cusp diagrams with heavy-quark eikonal lines.}
\end{center}
\label{1loop}
\end{figure}

We find the one-loop soft (cusp) anomalous dimension  
\beq
\Gamma_S^{(1)}=C_F \left[-\frac{(1+\beta^2)}{2\beta} 
\ln\left(\frac{1-\beta}{1+\beta}\right) -1\right]
\label{GammaS1}
\eeq
where $C_F=(N_c^2-1)/(2N_c)$ with $N_c=3$ the number of colors. 
This result can also be written in terms of the cusp angle \cite{KorRad} 
$\gamma=\ln[(1+\beta)/(1-\beta)]$, with $\coth\gamma=(1+\beta^2)/(2\beta)$, as
\beq
\Gamma_S^{(1)}=C_F (\gamma \coth\gamma -1) \, .
\eeq

\begin{figure}
\begin{center}
\includegraphics[width=9cm]{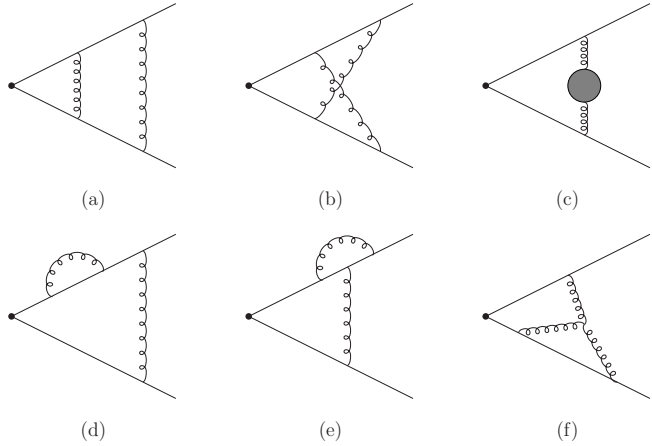}
\caption{Two-loop cusp vertex diagrams with heavy-quark eikonal lines.}
\end{center}
\label{2vloop}
\end{figure}

\begin{figure}
\begin{center}
\includegraphics[width=9cm]{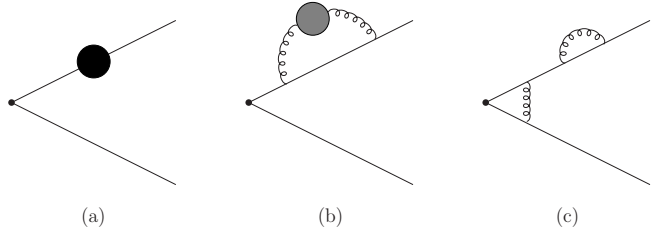}
\caption{Two-loop cusp self-energy diagrams with heavy-quark eikonal lines.}
\end{center}
\label{2sloop}
\end{figure}

\begin{figure}
\begin{center}
\includegraphics[width=9cm]{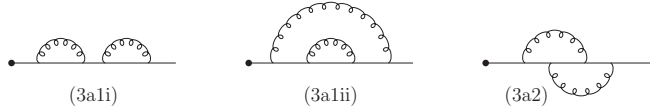}
\caption{Detail of the black blob of Fig. 3(a).}
\end{center}
\label{3a}
\end{figure} 

We now continue with the two-loop diagrams.  
In Fig. 2 we show graphs with vertex 
corrections and in Fig. 3 graphs with heavy-quark self-energy corrections. 
The dark blobs in Figs. 2(c) and 3(b) denote quark, gluon, and ghost loops.
The black blob in Fig. 3(a) denotes three kinds of corrections shown 
in Fig. 4.    
We do not show graphs with gluon loops involving four-gluon vertices and graphs 
involving three-gluon vertices with all three gluons attaching to a single eikonal line 
since such graphs have vanishing contributions.  
For each diagram we include the appropriate one-loop counterterms for the divergent 
subdiagrams. The calculations are challenging because of the presence of the heavy 
quark mass. Dimensionally-regularized integrals needed in the calculation 
are shown in Appendix A. Using the results in Appendix A,
the UV poles of the integrals for each diagram are provided in Appendix B.

Combining the kinematic results in Appendix B with color and symmetry factors, 
the contribution of the diagrams in Figs. 2 and 3 to the two-loop soft (cusp)  
anomalous dimension is  
\beqa
&&
C_F^2 \left[I_{2a}+I_{2b}+2 \, I_{2d}
+2 \, I_{2e}+I_{3a1}+I_{3a2}+I_{3c}\right]
\nonumber \\ 
{}&+&C_F \, C_A \left[-\frac{1}{2} I_{2b} 
+I_{2f} -I_{2cg}
-I_{2e}-I_{3bg}-\frac{1}{2} I_{3a2} \right]
+\frac{1}{2} C_F \left[I_{2cq}+I_{3bq}\right]  
\nonumber \\
&=&
-\frac{1}{2 \epsilon^2} \left(\Gamma_S^{(1)}\right)^2
+\frac{\beta_0}{4 \epsilon^2} \Gamma_S^{(1)}
-\frac{1}{2 \epsilon} \Gamma_S^{(2)}  
\label{S2}
\eeqa
where $I_k$ denotes the integral for diagram $k$, e.g. $I_{2d}$ 
is the integral for diagram 2(d). Also $I_{2cq}$ and $I_{3bq}$ denote 
the quark-loop contribution in Figs. 2(c) and 3(b), respectively, 
while $I_{2cg}$ and $I_{3bg}$ denote the gluon-loop plus ghost-loop 
contributions to the respective diagrams. $I_{3a1}$ denotes the sum 
of the graphs 3(a1i) and 3(a1ii) detailed in Fig. 4 while 
$I_{3a2}$ is the integral for the last graph in Fig. 4.
On the right-hand side of Eq. (\ref{S2})  in addition to the two-loop 
soft anomalous dimension, 
$\Gamma_S^{(2)}$, which appears in the coefficient of the  
$1/\epsilon$ pole, there also  appear terms from the exponentiation 
of the one-loop result and the running of the coupling 
which account for all the double poles of the graphs.
Here $\beta_0=(11/3) C_A-2n_f/3$, 
with $C_A=N_c$ and $n_f$ the number of light quark flavors. 
From Eq. (\ref{S2}) we solve for the two-loop soft (cusp) anomalous dimension:
\beq
\Gamma_S^{(2)}=\frac{K}{2} \, \Gamma_S^{(1)}
+C_F C_A M_{\beta}
\label{Gammas2}
\eeq
where 
\beqa
&& M_{\beta}=\frac{1}{2}+\frac{\zeta_2}{2}
+\frac{1}{2} \ln^2\left(\frac{1-\beta}{1+\beta}\right)
\nonumber \\ && \hspace{-10mm}
{}-\frac{(1+\beta^2)^2}{8 \beta^2} \left[\zeta_3
+\zeta_2 \ln\left(\frac{1-\beta}{1+\beta}\right)
+\frac{1}{3} \ln^3\left(\frac{1-\beta}{1+\beta}\right)
+\ln\left(\frac{1-\beta}{1+\beta}\right) 
{\rm Li}_2\left(\frac{(1-\beta)^2}{(1+\beta)^2}\right) 
-{\rm Li}_3\left(\frac{(1-\beta)^2}{(1+\beta)^2}\right)\right] 
\nonumber \\ &&  \hspace{-10mm}
{}-\frac{(1+\beta^2)}{4 \beta} \left[\zeta_2
-\zeta_2 \ln\left(\frac{1-\beta}{1+\beta}\right) 
+\ln^2\left(\frac{1-\beta}{1+\beta}\right)
-\frac{1}{3} \ln^3\left(\frac{1-\beta}{1+\beta}\right)
+2  \ln\left(\frac{1-\beta}{1+\beta}\right) 
\ln\left(\frac{(1+\beta)^2}{4 \beta}\right) \right. 
\nonumber \\ &&  \hspace{15mm} \left.
{}-{\rm Li}_2\left(\frac{(1-\beta)^2}{(1+\beta)^2}\right)\right]\, .
\label{Mbeta}
\eeqa
We have written the two-loop result $\Gamma_S^{(2)}$ 
in Eq. (\ref{Gammas2}) in the form of a term which is a multiple 
of the one-loop soft anomalous dimension $\Gamma_S^{(1)}$, Eq. (\ref{GammaS1}),
plus a set of additional terms which have been denoted as $M_{\beta}$.
Here $\zeta_2=\pi^2/6$ and $\zeta_3=1.2020569\cdots$.
The well-known two-loop constant $K$ \cite{JKLT} is given by 
$K=C_A (67/18-\zeta_2)-5n_f/9$. The color structure of $\Gamma_S^{(2)}$ 
involves only the factors $C_F C_A$ and $C_F n_f$.
Note that as $\beta \rightarrow 1$, $M_{\beta} \rightarrow (1-\zeta_3)/2$.

The result in Eq. (\ref{Gammas2}) can be written in terms 
of the cusp angle $\gamma$ as
\beqa
\Gamma_S^{(2)}&=&\frac{K}{2} \, \Gamma_S^{(1)}
+C_F C_A \left\{\frac{1}{2}+\frac{\zeta_2}{2}+\frac{\gamma^2}{2}
-\frac{1}{2}\coth^2\gamma\left[\zeta_3-\zeta_2\gamma-\frac{\gamma^3}{3}
-\gamma \, {\rm Li}_2\left(e^{-2\gamma}\right)
-{\rm Li}_3\left(e^{-2\gamma}\right)\right] \right.
\nonumber \\ && \hspace{25mm} \left.
{}-\frac{1}{2} \coth\gamma\left[\zeta_2+\zeta_2\gamma+\gamma^2
+\frac{\gamma^3}{3}+2\, \gamma \, \ln\left(1-e^{-2\gamma}\right)
-{\rm Li}_2\left(e^{-2\gamma}\right)\right] \right\},
\label{2lcusp}
\eeqa
and is in agreement, but in a simpler and more explicit form, with the result for 
the cusp anomalous dimension of Ref. \cite{KorRad}. 

\mysection{Two-loop soft anomalous dimension and NNLL resummation for $bg \rightarrow tW^-$ and $bg \rightarrow tH^-$}

\begin{figure}
\begin{center}
\includegraphics[width=10cm]{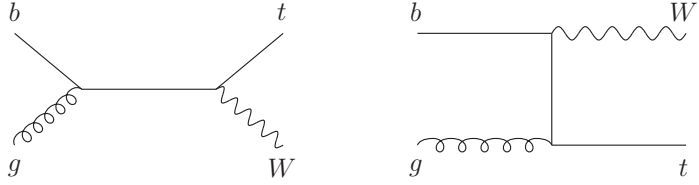}
\caption{Leading-order diagrams for $bg \rightarrow tW^-$.}
\label{tWborn}
\end{center}
\end{figure}

We now turn our attention to processes that involve a bottom quark, a gluon, and a top quark  
as the colored particles in the hard scattering, namely $tW^-$ and $tH^-$ production. 
The leading-order diagrams for $bg \rightarrow tW^-$ are shown in 
Fig. \ref{tWborn};  if one replaces the $W^-$ by an $H^-$ the graphs 
describe $bg \rightarrow tH^-$.
We treat the bottom quark as massless \cite{NKchiggs}.
In this section we calculate the two-loop soft anomalous 
dimension that will allow us to resum the soft-gluon contributions 
to NNLL accuracy.

\begin{figure}
\begin{center}
\includegraphics[width=9cm]{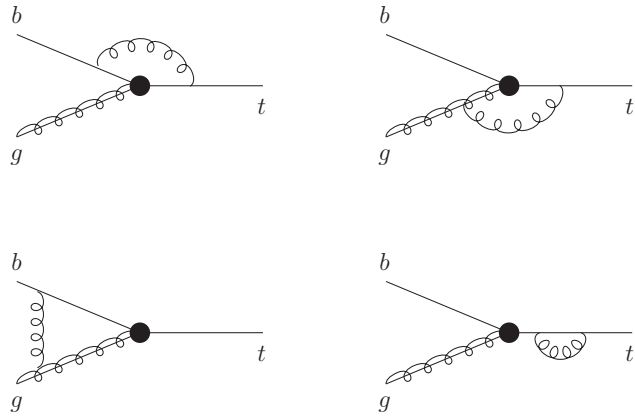}
\caption{One-loop eikonal diagrams with bottom quark-gluon-top quark  vertex.}
\label{tW1loop}
\end{center}
\end{figure}

In Fig. \ref{tW1loop} we show the one-loop eikonal diagrams for these
processes. Calculating the integrals associated with these diagrams we find
the one-loop soft anomalous dimension for $bg \rightarrow tW^-$:
\beq
\Gamma_{S,\, tW^-}^{(1)}=C_F \left[\ln\left(\frac{m_t^2-t}{m_t\sqrt{s}}\right)
-\frac{1}{2}\right] +\frac{C_A}{2} \ln\left(\frac{m_t^2-u}{m_t^2-t}\right)
\label{tW1l}
\eeq
where $s=(p_b+p_g)^2$, $t=(p_b-p_t)^2$, $u=(p_g-p_t)^2$, and $m_t$ is the 
top quark mass. The expression for $bg \rightarrow tH^-$ is identical. 
This result is slightly different from the result in Ref. \cite{NKst,NKchiggs} 
because the axial gauge was used in those papers, while the result in 
Eq. (\ref{tW1l}) is calculated in Feynman gauge. 
Of course these differences are compensated by other terms in the 
resummed formula and the  
final result for the cross section is independent of the choice of gauge.

\begin{figure}
\begin{center}
\includegraphics[width=8cm]{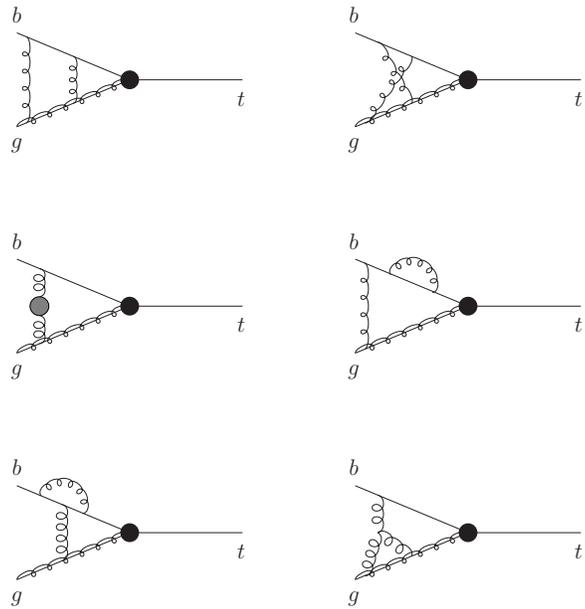}
\caption{Two-loop eikonal diagrams involving the bottom quark and gluon eikonal lines.}
\label{2lv1c}
\end{center}
\end{figure}

\begin{figure}
\begin{center}
\includegraphics[width=8cm]{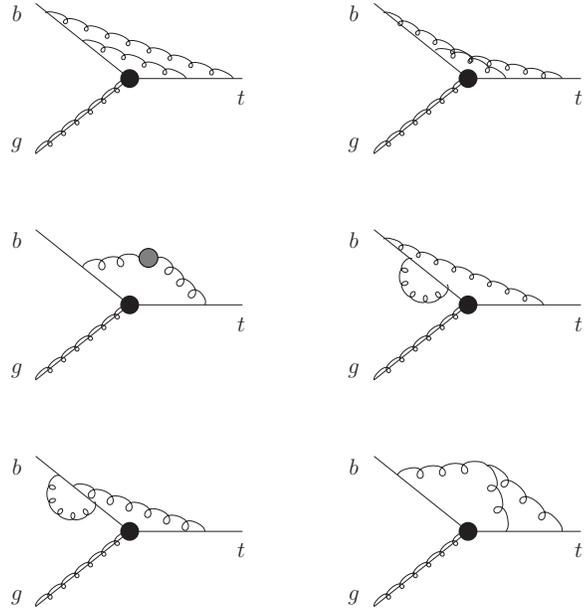}
\caption{Two-loop eikonal diagrams involving the bottom quark and top quark eikonal lines.}
\label{2lv12}
\end{center}
\end{figure}

\begin{figure}
\begin{center}
\includegraphics[width=8cm]{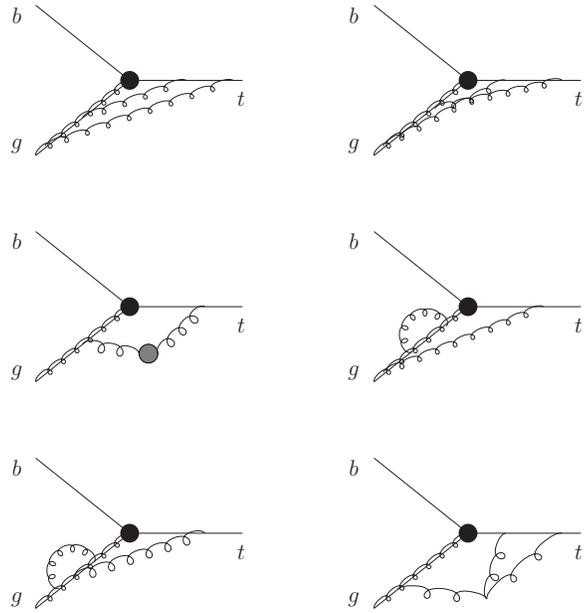}
\caption{Two-loop eikonal diagrams involving the gluon and top quark eikonal lines.}
\label{2lvc2}
\end{center}
\end{figure}

To find the two-loop soft anomalous dimension we calculate the diagrams shown in 
Figs. \ref{2lv1c}, \ref{2lv12}, \ref{2lvc2}, plus diagrams involving the top-quark 
self energy as in Figs. 3 and 4. Since they are three colored partons, with 
one of them massive,  we calculate diagrams that contribute to the cusp anomalous dimension 
for each pair of partons using the results in the previous section in the limit when one or both partons 
are massless. Note that diagrams that involve gluons attached to all three eikonal lines 
either vanish or do not contribute to the two-loop result \cite{ADS}, and hence we do not show them.
Combining the kinematic results for the integrals from Appendix B with color and symmetry factors we have
\beqa
&&
\frac{C_A^2}{4}\left[I_{2a}+2\, I_{2b}+2 \, I_{2cg}
+2 \, I_{2e}-2\, I_{2f}\right]_{bt}
+C_F^2 \left[I_{2a}+I_{2b}+2 \, I_{2d}
+2 \, I_{2e}+I_{3a}+I_{3c}\right]_{bt}
\nonumber \\ 
{}&+&C_F \, C_A \left[-I_{2a}-\frac{3}{2} I_{2b} 
-I_{2cg}-I_{2d}-2\, I_{2e}+I_{2f}-I_{3bg}-\frac{1}{2} I_{3a2}
-\frac{1}{2}I_{3c} \right]_{bt}
-\frac{C_A}{4} [I_{2cq}]_{bt}
\nonumber \\ 
{}&+&\frac{1}{2} C_F \left[I_{2cq}+I_{3bq}\right]_{bt}  
+\frac{C_A^2}{4}\left[I_{2a}-2 \, I_{2cg}
+2 \, I_{2d}+2\, I_{2f}\right]_{bg}
+\frac{C_A}{4} [I_{2cq}]_{bg}
\nonumber \\ 
&+&\frac{C_A^2}{4}\left[I_{2a}-2 \, I_{2cg}
+2 \, I_{2d}+2\, I_{2f}\right]_{gt}
+\frac{C_A}{4} [I_{2cq}]_{gt}+C_F C_A\frac{1}{2} [I_{3c}]_{gt}
+I_{3-{\rm line}}
\nonumber \\
&=&
-\frac{1}{2 \epsilon^2} \left(\Gamma_{S,\, tW^-}^{(1)}\right)^2
+\frac{\beta_0}{4 \epsilon^2} \Gamma_{S,\, tW^-}^{(1)}
-\frac{1}{2 \epsilon} \Gamma_{S,\, tW^-}^{(2)} 
\label{S2tW}
\eeqa
where $I_{3-{\rm line}}$ denotes the terms involving gluons attached to all three lines that 
do not contribute at two loops and $[I_{2d}]_{bt}$, for example, stands for the 2(d)-type diagram 
in Fig. 8 involving the $b$ and $t$ quarks.

We thus find the two-loop soft-anomalous dimension for $bg \rightarrow tW^-$
\beq
\Gamma_{S,\, tW^-}^{(2)}=\frac{K}{2} \Gamma_{S,\, tW^-}^{(1)}
+C_F C_A \frac{(1-\zeta_3)}{4}
\nonumber 
\eeq
where $\Gamma_{S, \, tW^-}^{(1)}$ is given in Eq. (\ref{tW1l}).
The result for $bg \rightarrow tH^-$ is the same.

With the two-loop soft-anomalous dimension at hand we are now ready to 
resum the soft-gluon corrections at NNLL accuracy. 
For $tW^-$  production the resummed partonic cross section in moment space 
(with $N$ the moment variable) is given by \cite{NKGS,NKst,NKsch,NNNLO} 
\beqa
{\hat{\sigma}}^{res}(N) &=&   
\exp\left[ E_q(N_q)+E_g(N_g)\right] \;  
\exp \left[ 2 \int_{\mu_F}^{\sqrt{s}} \frac{d\mu}{\mu}\;
\left(\gamma_{q/q}\left({\tilde N}_q, \alpha_s(\mu)\right)
+\gamma_{g/g}\left({\tilde N}_g, \alpha_s(\mu)\right)\right)
\right] \;
\nonumber\\ && \hspace{-10mm} \times \,
H^{bg \rightarrow tW^-}
\left(\alpha_s(\sqrt{s})\right) \;
S^{bg \rightarrow tW^-}\left(\alpha_s(\sqrt{s}/{\tilde N'})
\right) \; \exp \left[2\int_{\sqrt{s}}^{{\sqrt{s}}/{\tilde N'}} 
\frac{d\mu}{\mu}\; \Gamma_{S,\, tW^-}
\left(\alpha_s(\mu)\right)\right]  
\label{resHS}
\eeqa
and similarly for $tH^-$ production.

The first exponent \cite{GS87,CT89} in the above expression 
resums soft and collinear corrections from the incoming partons
\beq
E_i(N_i)=
\int^1_0 dz \frac{z^{N_i-1}-1}{1-z}\;
\left \{\int_1^{(1-z)^2} \frac{d\lambda}{\lambda}
A_i\left(\alpha_s(\lambda s)\right)
+D_i\left[\alpha_s((1-z)^2 s)\right]\right\} 
\label{Eexp}
\eeq
where $i$ stands for the incoming bottom quark ($i=q$) or the 
incoming gluon ($i=g$).
Here $N_q=N[(m_W^2-u)/m_t^2]$ and 
$N_g=N[(m_W^2-t)/m_t^2]$ where $m_W$ is the $W$-boson mass. 
The quantity $A_i$ has a perturbative expansion, 
$A_i=\sum_n (\alpha_s/\pi)^n A_i^{(n)}$. 
Here 
$A_q^{(1)}=C_F$ and $A_q^{(2)}=C_F K/2$, while  
$A_g^{(1)}=C_A$ and $A_g^{(2)}=C_A K/2$. 

Also $D_i=\sum_n (\alpha_s/\pi)^n D_i^{(n)}$, 
with $D_q^{(1)}=D_g^{(1)}=0$, and \cite{CLS97}
\beq
D_q^{(2)}=C_F C_A \left(-\frac{101}{54}+\frac{11}{6} \zeta_2
+\frac{7}{4}\zeta_3\right)
+C_F n_f \left(\frac{7}{27}-\frac{\zeta_2}{3}\right) 
\eeq
and $D_g^{(2)}=(C_A/C_F) D_q^{(2)}$.

In the third exponent $\gamma_{i/i}$ is the moment-space 
anomalous dimension of the ${\overline {\rm MS}}$ 
parton density $\phi_{i/i}$ and it controls the factorization scale, $\mu_F$, 
dependence of the cross section. 
We have $\gamma_{i/i}=-A_i \ln {\tilde N}_i +\gamma_i$ where $A_i$ was 
defined above, ${\tilde N}_i=N_i e^{\gamma_E}$ 
with $\gamma_E$ the Euler constant, and the parton anomalous dimension
$\gamma_i=\sum_n (\alpha_s/\pi)^n \gamma_i^{(n)}$
where  $\gamma_q^{(1)}=3C_F/4$ and $\gamma_g^{(1)}=\beta_0/4$.

$H^{bg \rightarrow tW}$ is the hard-scattering function 
while $S^{bg \rightarrow tW}$ is the 
soft function describing noncollinear soft gluon emission \cite{NKGS,NKst}.
The evolution of the soft function is controlled by the soft anomalous 
dimension $\Gamma_{S,\, tW^-}$.
Here ${\tilde N'}={\tilde N}(s/m_t^2)$ with ${\tilde N}=N e^{\gamma_E}$.

For $tH^-$ production the resummed formula is essentially the same. 
The only difference, apart from the obvious use of the appropriate 
hard-scattering function for this process, is the definition 
of $N_q$ and $N_g$. In this case, $N_q=N[(m_{H^-}^2-u)/m_{H^-}^2]$ and 
$N_g=N[(m_{H^-}^2-t)/m_{H^-}^2]$ where $m_{H^-}$ is the charged Higgs mass. 

The resummed cross section, Eq. (\ref{resHS}),  
can be expanded in the strong coupling, $\alpha_s$, 
and inverted to momentum space, thus providing fixed-order results 
for the soft-gluon corrections.
The NLO expansion of the resummed cross section after inversion to 
momentum space is 
\beq
{\hat{\sigma}}^{(1)} = \sigma^B \frac{\alpha_s(\mu_R)}{\pi}
\left\{c_3\, {\cal D}_1(s_4) + c_2\,  {\cal D}_0(s_4) \right\} \, ,
\label{NLOmaster}
\eeq
where $\sigma^B$ is the Born term for the process and $\mu_R$ is the 
renormalization scale. We use the notation
${\cal D}_k(s_4)=[\ln^k(s_4/m_t^2)/s_4]_+$ in $tW^-$ production 
and ${\cal D}_k(s_4)=[\ln^k(s_4/m_{H^-}^2)/s_4]_+$ in $tH^-$ production 
for the plus distributions 
involving logarithms of a kinematical variable $s_4$ that measures distance 
from threshold ($s_4=0$ at threshold). For $bg \rightarrow tW^-$, 
$s_4=s+t+u-m_t^2-m_W^2$, while for $bg \rightarrow tH^-$, 
$s_4=s+t+u-m_t^2-m_{H^-}^2$. 
The coefficient of the leading term 
is
\beq
c_3= 2(A_q^{(1)}+A_g^{(1)})\, .
\label{c3}
\eeq 
The coefficient of the next-to-leading term, $c_2$, can be written as
 $c_2=c_2^{\mu}+T_2$, with $c_2^{\mu}$ denoting the terms involving logarithms 
of the scale and $T_2$ denoting the scale-independent terms.  
For $bg \rightarrow tW^-$
\beq
c_2^{\mu}=-(A_q^{(1)}+A_g^{(1)}) \ln\left(\frac{\mu_F^2}{m_t^2}\right)
\eeq
and  
\beq
T_2=-2 \, A_q^{(1)} \, \ln\left(\frac{m_W^2-u}{m_t^2}\right)
-2 \, A_g^{(1)} \, \ln\left(\frac{m_W^2-t}{m_t^2}\right)
-(A_q^{(1)}+A_g^{(1)}) \ln\left(\frac{m_t^2}{s}\right)
+2 \Gamma_{S,\, tW^-}^{(1)}\, .
\label{c2n}
\eeq
For $bg \rightarrow tH^-$ replace both $m_W$ and $m_t$ in the above two 
equations by $m_{H^-}$.

As discussed in \cite{NKst,NKchiggs} the expansion can also determine 
the terms involving logarithms of the factorization and renormalization 
scales in the coefficient, 
$c_1$, of the $\delta(s_4)$ terms. If we denote these terms as
$c_1^{\mu}$, then for $tW^-$ production 
\beq
c_1^{\mu}=\left[A_q^{(1)}\, \ln\left(\frac{m_W^2-u}{m_t^2}\right) 
+A_g^{(1)}\, \ln\left(\frac{m_W^2-t}{m_t^2}\right)
-\gamma_q^{(1)}-\gamma_g^{(1)}\right]\ln\left(\frac{\mu_F^2}{m_t^2}\right) 
+\frac{\beta_0}{4} \ln\left(\frac{\mu_R^2}{m_t^2}\right)\, ,
\label{c1mu}
\eeq
while for $bg \rightarrow tH^-$ replace both $m_W$ and $m_t$ in the above  
equation by $m_{H^-}$.
The full virtual terms are not derivable from resummation, which addresses 
soft-gluon contributions, but can be taken from the complete NLO calculation.

The NNLO expansion of the resummed cross section 
for $bg \rightarrow tW^-$ after inversion to 
momentum space is
\beqa
{\hat{\sigma}}^{(2)}&=&\sigma^B \frac{\alpha_s^2(\mu_R)}{\pi^2}
\left\{\frac{1}{2}c_3^2\, {\cal D}_3(s_4) + 
\left[\frac{3}{2}c_3 c_2-\frac{\beta_0}{4} c_3\right]  {\cal D}_2(s_4) \right.
\nonumber \\ && \hspace{-5mm} 
{}+\left[c_3 c_1+c_2^2-\zeta_2 c_3^2
-\frac{\beta_0}{2} T_2+\frac{\beta_0}{4} c_3 
\ln\left(\frac{\mu_R^2}{m_t^2}\right)+2 A_q^{(2)}+2 A_g^{(2)}\right] {\cal D}_1(s_4)
\nonumber \\ && \hspace{-5mm}  
{}+\left[c_2 c_1-\zeta_2 c_3 c_2+\zeta_3 c_3^2
+\frac{\beta_0}{4} c_2 \ln\left(\frac{\mu_R^2}{s}\right) \right. 
-\frac{\beta_0}{2} A_q^{(1)} \ln^2\left(\frac{m_W^2-u}{m_t^2}\right)
-\frac{\beta_0}{2} A_g^{(1)} \ln^2\left(\frac{m_W^2-t}{m_t^2}\right)
\nonumber \\ && \hspace{-5mm} \quad \quad  
{}-2 A_q^{(2)} \ln\left(\frac{m_W^2-u}{m_t^2}\right)
-2 A_g^{(2)} \ln\left(\frac{m_W^2-t}{m_t^2}\right)+D_q^{(2)}+D_g^{(2)}
\nonumber \\ && \hspace{-5mm}  \quad \quad \left. \left. 
{}+\frac{\beta_0}{8}  (A_q^{(1)}+A_g^{(1)}) \ln^2\left(\frac{\mu_F^2}{s}\right) 
-(A_q^{(2)}+A_g^{(2)}) \ln\left(\frac{\mu_F^2}{s}\right)
+2 \Gamma_{S,\, tW^-}^{(2)}\right]  {\cal D}_0(s_4) \right\} \, .
\label{NNLOapprox}
\eeqa
For $bg \rightarrow tH^-$ again replace both $m_W$ and $m_t$ by $m_{H^-}$ 
in the above equation.
It is important to note that all NNLO soft-gluon corrections are derived from the 
NNLL resummed cross section, i.e. the coefficients of all powers 
of logarithms in $s_4$ are given in Eq. (\ref{NNLOapprox}), from ${\cal D}_3(s_4)$ down 
to ${\cal D}_0(s_4)$. In Ref. \cite{NKst} and \cite{NKchiggs,NKchiggs2}, 
where NLL accuracy 
was attained, only the coefficients of ${\cal D}_3(s_4)$ and ${\cal D}_2(s_4)$ 
were fully determined. 
Thus, at NNLL accuracy the theoretical improvement over NLL
is significant. As discussed in \cite{NKst,NKchiggs2} additional 
$\delta(s_4)$ terms involving the factorization 
and renormalization scales are also computed.

\mysection{NNLO approximate cross sections for $tW^-$ and $tH^-$ production 
at the LHC}

\begin{table}[htb]
\begin{center}
\begin{tabular}{|c|c|c|c|} \hline
\multicolumn{4}{|c|}{NNLO approx (NNLL) $tW^-$ cross section (pb)} \\ \hline
$m_t$ (GeV) & LHC 7 TeV & LHC 10 TeV & LHC 14 TeV \\ \hline
170 & 8.24 & 20.3 & 43.6 \\ \hline 
171 & 8.09 & 20.0 & 43.0 \\ \hline 
172 & 7.94 & 19.7 & 42.4\\ \hline 
173 & 7.80 & 19.4 & 41.8 \\ \hline 
174 & 7.66 & 19.1 & 41.2 \\ \hline 
175 & 7.53 & 18.7 & 40.6 \\ \hline 
\end{tabular}
\caption[]{The $bg \rightarrow tW^-$ production cross section in pb 
in $pp$ collisions at the LHC with $\sqrt{S}=7$ TeV, 10 TeV, 
and 14 TeV, with 
$\mu=m_t$ and using the MSTW2008 NNLO pdf \cite{MSTW2008}.
The approximate NNLO results are shown at NNLL accuracy.}
\label{table1}
\end{center}
\end{table}

We now use the results of the previous section to calculate approximate 
NNLO cross sections for $bg \rightarrow tW^-$ and 
$bg\rightarrow tH^-$ at the LHC.

We begin with $tW^-$ production.
As has been shown in \cite{NKst,NKstlhc} the NLO expansion of the
resummed cross section approximates well the complete NLO result 
for both Tevatron and LHC energies. In fact when damping factors are used to 
limit the soft-gluon contributions far away from threshold, as was also used 
for $t{\bar t}$ production \cite{NKRV} and $s$-channel single-top production 
\cite{NKsch}, then the approximation is excellent. This shows that soft-gluon 
corrections are dominant for this process.

In Table 1 we provide numerical values for the $tW^-$ 
cross section at the LHC for energies of 7, 10, and 14 TeV and a range 
of top quark masses 
from 170 to 175 GeV. The NNLO approximate corrections increase the NLO cross 
section by $\sim 8$\%. We note that the cross section for 
${\bar b} g \rightarrow {\bar t} W^+$ is identical.

At 7 TeV with $m_t=173$ GeV the approximate NNLO cross section from NNLL 
resummation is 
\beq
\sigma^{\rm NNLOapprox}_{tW^-}(m_t=173 \, {\rm GeV}, \, 7\, {\rm TeV})
=7.8 \pm 0.2 {}^{+0.5}_{-0.6} \; {\rm pb}\, .
\eeq
The first uncertainty is from scale variation between $m_t/2$ and $2m_t$ 
and the second is from the MSTW2008 NNLO pdf at 90\% C.L.
At 10 TeV, again with $m_t=173$ GeV, the cross section is 
$19.4 \pm 0.5 {}^{+1.0}_{-1.1}$ pb, and at 14 TeV we find 
$41.8 \pm 1.0 {}^{+1.5}_{-2.4}$ pb.

\begin{figure}
\begin{center}
\includegraphics[width=11cm]{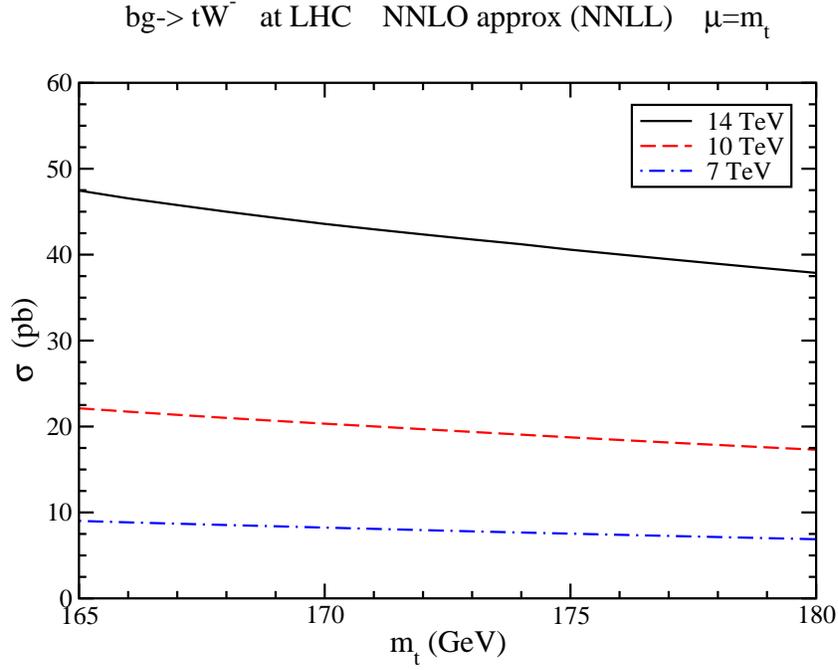}
\caption{The cross section for $tW^-$ production 
at the LHC with $\sqrt{S}=7$ TeV, 10 TeV, and 14 TeV, and MSTW2008 NNLO pdf.}
\label{LHCtW}
\end{center}
\end{figure}

In Fig. \ref{LHCtW} we plot the $bg \rightarrow tW^-$ NNLO approximate cross 
section from NNLL resummation at the LHC versus top quark mass 
for energies of 7, 10, and 14 TeV. 

\begin{figure}
\begin{center}
\includegraphics[width=11cm]{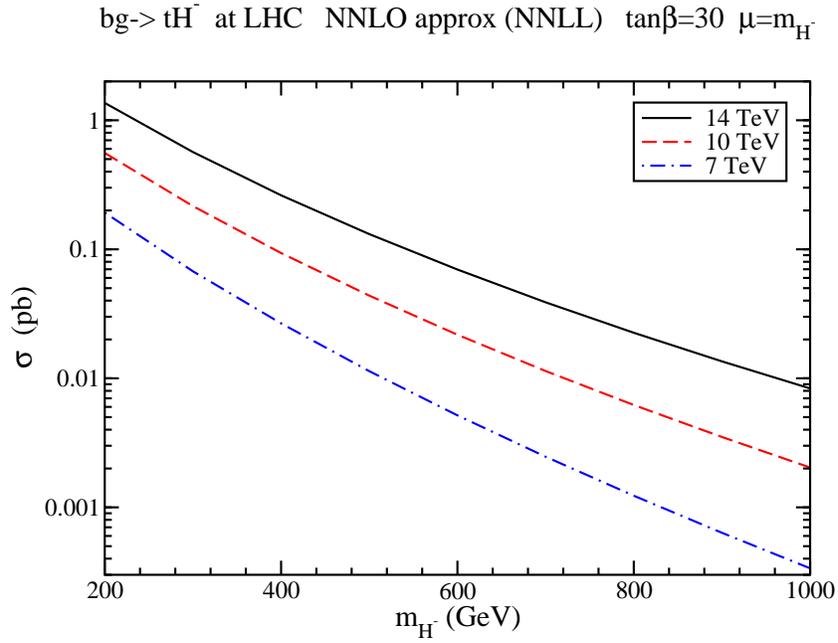}
\caption{The cross section for $tH^-$ production 
at the LHC with $\sqrt{S}=7$ TeV, 10 TeV, and 14 TeV, and MSTW2008 NNLO pdf.}
\label{LHCtH}
\end{center}
\end{figure}

Next we consider the process $bg \rightarrow tH^-$. 
The ratio of the vacuum expectation values, $v_2,v_1$ for the two Higgs
doublets is $\tan \beta=v_2/v_1$, and the value of the cross section depends 
on the choice of this undetermined parameter. However, the overall percentage
enhancement of the cross section from the higher-order soft-gluon corrections
is independent of the value of $\tan\beta$.

In Fig. \ref{LHCtH} we plot the $bg \rightarrow tH^-$ NNLO approximate cross 
section from NNLL resummation at the LHC versus charged Higgs mass 
for energies of 7, 10, and 14 TeV, using a value of $\tan\beta=30$.
The NNLO approximate corrections increase the NLO cross section by 
$\sim 15$\% to $\sim 20$\% for the range of charged Higgs masses shown.
We note that the cross section for ${\bar b}g \rightarrow {\bar t} H^+$ 
is identical (assuming the underlying model is $CP$ conserving).

\mysection{Conclusion}

The cross sections for associated production of a single top quark with 
a $W$ boson or with a charged Higgs boson receive large contributions 
from soft gluon corrections. These contributions were resummed in this 
paper to NNLL accuracy, thus extending previous NLL results. 
Attaining this accuracy requires the calculation of two-loop soft anomalous 
dimensions from the UV poles of dimensionally regularized integrals 
of two-loop eikonal diagrams. First the two-loop cusp anomalous dimension 
was calculated, which is an essential ingredient to all NNLL resummations 
for QCD processes, and then the result was used to calculate the two-loop 
soft anomalous dimensions for $bg \rightarrow tW^-$ and $bg \rightarrow tH^-$. 
From the NNLL resummed formula approximate NNLO cross sections were derived 
and numerical predictions made for $tW^-$ and $tH^-$ production at LHC 
energies. These approximate NNLO corrections enhance the NLO cross section 
for $tW^-$ production by $\sim 8$\% and for  $tH^-$ production  
by $\sim 15$\% to $\sim 20$\%.

\mysection*{Acknowledgements}
This work was supported by the National Science Foundation under 
Grant No. PHY 0855421. 

\renewcommand{\theequation}{A.\arabic{equation}}
\setcounter{equation}{0}
\section*{Appendix A: Dimensionally regularized eikonal integrals}

We list results for several dimensionally regularized integrals needed 
in the calculation of the two-loop soft anomalous dimension.
 
\beq
\int\frac{d^n k}{k^2 \, v_i\cdot k \, v_j\cdot k}=\frac{i}{\epsilon}
(-1)^{-1-\frac{\epsilon}{2}} \pi^{2-\frac{\epsilon}{2}} 
2^{3+3\frac{\epsilon}{2}}
\Gamma\left(1+\frac{\epsilon}{2}\right)\, {}_2F_1\left(\frac{1}{2},
1+\frac{\epsilon}{2};\frac{3}{2};\beta^2\right)
\label{A1}
\eeq
where ${}_2F_1$ is the Gauss hypergeometric function.

\beq
\int\frac{d^n k}{k^2 \, (v_i\cdot k)^2}
=\frac{i}{\epsilon}\,
(-1)^{1-\frac{\epsilon}{2}} \, \pi^{2-\frac{\epsilon}{2}} \,  
2^{3+3\frac{\epsilon}{2}} 
(1-\beta^2)^{-1-\frac{\epsilon}{2}} 
\Gamma\left(1+\frac{\epsilon}{2}\right)
\label{A2}
\eeq

\beqa
&& \hspace{-10mm}
\int\frac{d^n k}{(k^2)^{1+\frac{\epsilon}{2}} \, v_i\cdot k \, v_j\cdot k}
=\frac{i}{\epsilon^2} \, \frac{(-1)^{1-\epsilon}}{\beta} \, 2^{2\epsilon} 
\pi^{2-\frac{\epsilon}{2}} \, \Gamma(1+\epsilon) 
\frac{1}{\Gamma\left(1+\frac{\epsilon}{2}\right)} 
\nonumber\\ &&  \hspace{-5mm} \times 
\left[(1-\beta)^{-\epsilon} {}_2F_1\left(-\epsilon,
1+\epsilon;1-\epsilon;\frac{1-\beta}{2}\right)
{}-(1+\beta)^{-\epsilon} {}_2F_1\left(-\epsilon,
1+\epsilon;1-\epsilon;\frac{1+\beta}{2}\right)\right]
\label{A3}
\eeqa

\beqa
&& \hspace{-9mm}
\int\frac{d^n k}{k^2 \, (v_i\cdot k)^{1+\epsilon} \, v_j\cdot k}
=\frac{i \pi^{2-\frac{\epsilon}{2}}}{\epsilon (1+\epsilon)} 
2^{2+\frac{9\epsilon}{2}} (-1)^{-1-\frac{3\epsilon}{2}}
(1-\beta^2)^{-1-\frac{3\epsilon}{2}} \, 
\Gamma\left(1+\frac{3\epsilon}{2}\right) \frac{1}{\Gamma(1+\epsilon)}
\nonumber\\ &&  \times 
F_1[1+\epsilon;1+\frac{3\epsilon}{2},1+\frac{3\epsilon}{2};2+\epsilon;
\frac{2\beta}{1+\beta},\frac{-2\beta}{1-\beta}] 
\label{A4}
\eeqa
where $F_1$ is the Appell hypergeometric function.

\beqa
&& \hspace{-7mm}
\int\frac{d^n k_2}{k_2^2 \, \left[v_i\cdot (k_1+k_2)\right]^2}
=\frac{i}{\epsilon } \, \frac{(-1)^{-1+\frac{\epsilon}{2}}}{(1+\epsilon)} \, 
2^{4-\frac{\epsilon}{2}} \, \pi^{\frac{3-\epsilon}{2}} 
(1-\beta^2)^{-1+\frac{\epsilon}{2}} \, (v_i \cdot k_1)^{-\epsilon} \, 
\nonumber\\ && \quad \times 
\Gamma\left(1+\frac{\epsilon}{2}\right) \,
\Gamma\left(1-\frac{\epsilon}{2}\right) \, 
\Gamma\left(\frac{3+\epsilon}{2}\right)  \, 
\label{A5}
\eeqa

\beq
\int\frac{d^n k}{k^2 \, (v_i\cdot k)^{2+\epsilon}}
=\frac{i \pi^{2-\frac{\epsilon}{2}}}{\epsilon (1+\epsilon)} 
2^{2+\frac{9\epsilon}{2}} (-1)^{-1-\frac{3\epsilon}{2}}
(1-\beta^2)^{-1-\frac{3\epsilon}{2}} \, 
\Gamma\left(1+\frac{3\epsilon}{2}\right) \frac{1}{\Gamma(1+\epsilon)} \, 
\label{A6}
\eeq

\beqa
&& \hspace{-7mm}
\int\frac{d^n k_1}{k_1^2 \, v_i\cdot k_1 \, v_i\cdot (k_1+k_2)}
=\frac{i}{\epsilon} \, (-1)^{\frac{\epsilon}{2}} \, 
2^{2-\frac{\epsilon}{2}} \, \pi^{\frac{3-\epsilon}{2}} 
(v_i \cdot k_2)^{-\epsilon} \, (1-\beta^2)^{-1+\frac{\epsilon}{2}} \,  
\nonumber\\ && \quad \quad \times
\Gamma\left(1+\frac{\epsilon}{2}\right)\, 
\Gamma\left(1-\frac{\epsilon}{2}\right) \, 
\Gamma\left(\frac{\epsilon-1}{2}\right)  \, 
\label{A7}
\eeqa

\beqa
&& \hspace{-7mm}
\int\frac{d^n k_2}{k_2^2 \, v_i\cdot k_2 \, [v_i\cdot (k_1+k_2)]^2}
=\frac{i}{1-\epsilon} \, (-1)^{1+\frac{\epsilon}{2}} 
2^{3-\frac{3\epsilon}{2}} \, \pi^{2-\frac{\epsilon}{2}} \, 
(v_i \cdot k_1)^{-1-\epsilon} \, 
\nonumber\\ && \quad \quad \times 
(1-\beta^2)^{-1+\frac{\epsilon}{2}} \, 
\Gamma\left(1-\frac{\epsilon}{2}\right) \, \Gamma(1+\epsilon)  \, 
\label{A8}
\eeqa

\beq
\int\frac{d^n k}{(k^2)^{1+\frac{\epsilon}{2}} \, (v_i\cdot k)^2}
=\frac{i}{\epsilon} \, (-1)^{-1-\epsilon} \, 
2^{2+3\epsilon} \, \pi^{2-\frac{\epsilon}{2}} 
(1-\beta^2)^{-1-\epsilon} \,  \Gamma(1+\epsilon) \, 
\frac{1}{\Gamma\left(1+\frac{\epsilon}{2}\right)}  \, 
\label{A9}
\eeq

\renewcommand{\theequation}{B.\arabic{equation}}
\setcounter{equation}{0}
\section*{Appendix B: UV poles of the integrals for eikonal one-loop and two-loop diagrams for the soft (cusp) anomalous dimension}

Here we present the UV poles of the integrals for the one-loop eikonal 
diagrams in Fig. 1 and the two-loop eikonal diagrams in Figs. 2 and 3.

First we list the integrals for the one-loop diagrams

\beq
I_{1a}=\frac{(1+\beta^2)}{2\, \beta}
\frac{1}{\epsilon} \ln\left(\frac{1-\beta}{1+\beta}\right) 
\label{I1aUV}
\eeq
and
\beq
I_{1b}=\frac{1}{\epsilon} 
\label{I1bUV}
\eeq

Then we list the integrals for the two-loop diagrams:  

\beq
I_{2a}+I_{2b}=\frac{(1+\beta^2)^2}{8 \, \beta^2} 
\frac{(-1)}{\epsilon^2} 
\ln^2\left(\frac{1-\beta}{1+\beta}\right) \, .
\eeq

\beq
I_{2b}=\frac{(1+\beta^2)^2}{8\beta^2} \frac{1}{\epsilon} 
\left\{-\frac{1}{3}\ln^3\left(\frac{1-\beta}{1+\beta}\right)
-\ln\left(\frac{1-\beta}{1+\beta}\right)
\left[{\rm Li}_2\left(\frac{(1-\beta)^2}{(1+\beta)^2}\right)
+\zeta_2\right] 
+{\rm Li}_3\left(\frac{(1-\beta)^2}{(1+\beta)^2}\right)-\zeta_3 \right\}
\nonumber \\
\eeq

\beq
I_{2cq}=n_f \frac{(1+\beta^2)}{6 \, \beta}
\left[\frac{1}{\epsilon^2}-\frac{5}{6 \, \epsilon}\right] 
\ln\left(\frac{1-\beta}{1+\beta}\right)
\eeq

\beq
I_{2cg}=\frac{5}{24} \frac{(1+\beta^2)}{\beta}  
\left[\frac{1}{\epsilon^2}-\frac{31}{30\, \epsilon}\right] 
\ln\left(\frac{1-\beta}{1+\beta}\right)
\eeq

\beqa
I_{2d}&=&\frac{(1+\beta^2)}{4\beta}
\left\{-\frac{1}{\epsilon^2} \ln\left(\frac{1-\beta}{1+\beta}\right)
+\frac{1}{\epsilon} \left[\ln\left(\frac{1-\beta}{1+\beta}\right)
+\frac{1}{2}\ln^2\left(\frac{1-\beta}{1+\beta}\right)\right. \right.
\nonumber \\ &&  \left. \left.
{}+\ln\left(\frac{1-\beta}{1+\beta}\right) 
\ln\left(\frac{(1+\beta)^2}{4\beta}\right)
-\frac{1}{2}{\rm Li}_2\left(\frac{(1-\beta)^2}{(1+\beta)^2}\right)
+\frac{\zeta_2}{2}\right] \right\} 
\eeqa

\beq
I_{2e}=-I_{2d}
\eeq

\beqa
I_{2f}=\frac{1}{\epsilon} \left\{-\frac{1}{4} \left[2 \zeta_2+\ln^2\left(\frac{1
-\beta}{1+\beta}\right)\right] 
\left[\frac{(1+\beta^2)}{2 \beta} \ln\left(\frac{1-\beta}{1+\beta}\right)+1\right]
+\frac{(1+\beta^2)}{12 \beta} \ln^3\left(\frac{1-\beta}{1+\beta}\right) \right\}
\eeqa

\beq
I_{3a1}=-\frac{3}{2 \epsilon^2}+\frac{1}{2\epsilon},
\eeq

\beq
I_{3a2}=\frac{1}{\epsilon^2}-\frac{1}{2 \epsilon}
\eeq

\beq
I_{3bq}=\frac{n_f}{3} \left[\frac{1}{\epsilon^2}-\frac{5}{6 \epsilon}\right]
\eeq

\beq
I_{3bg}=\frac{5}{12}\left[\frac{1}{\epsilon^2}-\frac{31}{30 \epsilon}\right] 
\eeq

\beq
I_{3c}= -\frac{1}{\epsilon^2} \frac{(1+\beta^2)}{2\beta} 
\ln[(1-\beta)/(1+\beta)] 
\eeq

In terms of the cusp angle 
\beq
\gamma=\ln\left(\frac{v_i \cdot v_j+\sqrt{(v_i \cdot v_j)^2-v_i^2 v_j^2}}{\sqrt{v_i^2 v_j^2}}\right)
\eeq
and 
\beq
\coth\gamma=\frac{v_i \cdot v_j}{\sqrt{(v_i \cdot v_j)^2-v_i^2 v_j^2}}
\eeq
the previous results for the integrals can be written as
\beq
I_{1a}=-\frac{1}{\epsilon}\gamma\coth\gamma
\eeq

\beq
I_{2a}+I_{2b}=-\frac{1}{2\epsilon^2} \gamma^2 \coth^2\gamma
\eeq

\beq
I_{2b}=\frac{1}{2\epsilon}\coth^2\gamma\left\{\gamma\left[{\rm Li}_2\left(e^{-2\gamma}\right)+\zeta_2\right]+\frac{\gamma^3}{3}+{\rm Li}_3\left(e^{-2\gamma}\right)-\zeta_3\right\}
\eeq

\beq
I_{2cq}=\frac{n_f}{3}\gamma\coth\gamma\left[-\frac{1}{\epsilon^2}+\frac{5}{6\epsilon}\right]
\eeq

\beq
I_{2cg}=\left[-\frac{5}{12\epsilon^2}+\frac{31}{72\epsilon}\right]\gamma\coth\gamma
\eeq

\beq
I_{2d}=\frac{1}{2}\coth\gamma\left\{\frac{1}{\epsilon^2}\gamma+\frac{1}{\epsilon}\left[\frac{\gamma^2}{2}-\gamma+\gamma\ln\left(1-e^{-2\gamma}\right)-\frac{1}{2}{\rm Li}_2\left(e^{-2\gamma}\right)+\frac{\zeta_2}{2}\right]\right\}
\eeq

\beq
I_{2f}=\frac{1}{\epsilon}\left\{-\frac{1}{4}\left[2\zeta_2+\gamma^2\right]\left[-\gamma\coth\gamma+1\right]-\frac{1}{6}\gamma^3\coth\gamma\right\}
\eeq

\beq
I_{3c}=\frac{1}{\epsilon^2}\gamma\coth\gamma
\eeq

The above expressions simplify when one of the quarks is massless. 
In that case $\coth\gamma=1$
and
\beq
\gamma=\ln\left(\frac{2 v_i \cdot v_j}{\sqrt{v_i^2 v_j^2}}\right)
\eeq
The integrals listed before then take simpler forms:
\beq
I_{1a}=-\frac{1}{\epsilon}\gamma
\eeq

\beq
I_{2a}+I_{2b}=-\frac{1}{2\epsilon^2} \gamma^2
\eeq

\beq
I_{2b}=\frac{1}{2\epsilon}\left[\frac{\gamma^3}{3}+\zeta_2\gamma-\zeta_3\right\}
\eeq

\beq
I_{2cq}=\frac{n_f}{3}\gamma\left[-\frac{1}{\epsilon^2}+\frac{5}{6\epsilon}\right]
\eeq

\beq
I_{2cg}=\left[-\frac{5}{12\epsilon^2}+\frac{31}{72\epsilon}\right]\gamma
\eeq

\beq
I_{2d}=\frac{1}{2\epsilon^2}\gamma+\frac{1}{2\epsilon}\left[\frac{\gamma^2}{2}-\gamma+\frac{\zeta_2}{2}\right]
\eeq

\beq
I_{2f}=\frac{1}{\epsilon}\left[\frac{1}{12}\gamma^3-\frac{1}{4}\gamma^2+\frac{\zeta_2}{2}\gamma-\frac{\zeta_2}{2}\right]
\eeq

\beq
I_{3c}=\frac{1}{\epsilon^2}\gamma
\eeq

\end{document}